# INHOMOGENEOUS TRAVELLING-WAVE ACCELERATING SECTIONS AND WKB APPROACH

*M.I. Ayzatsky*
*National Science Center Kharkov Institute of Physics and Technology (NSC KIPT), 610108, Kharkov, Ukraine*
*E-mail: aizatsky@kipt.kharkov.ua*

The paper presents the results of a study of the possibility of using the WKB approach to describe Inhomogeneous Travelling-Wave Accelerating Sections (ITWAS). This possibility not only simplifies the calculation, but also allows the use of simpler physical models of transient processes. Using the traveling wave concept simplifies the understanding of pulsed-excited ITWAS transients and the development of methods to mitigate their effect on beam parameters.
PACS: 29.20.−c ; 84.40.Az

## 1. INRODUCTION

Inhomogeneous Travelling-Wave Accelerating Sections (ITWAS) have been (and are) the workhorse of accelerating technology for more than half a century. Several thousand different sections were manufactured and used in linacs. Only one linac (SLAC) included 960 sections [1]. ITWASs are in fact chains of coupled resonators connected with two external waveguides. This apparent simplicity of structure is very deceiving. The reason is that the homogeneous periodic waveguide has the infinite number of eigen waves, most of which do not propagate (evanescent waves). Any inhomogeneity leads to the appearance such fields that decay exponentially from the interface at which they are formed. In ITWAS there are many small discontinuities with small field disturbance. Developing an electrodynamic model that combines propagation and evanescence is not an easy task. This is proved by the fact that before "computer age", as we know, only one mathematical model that could rigorously describe characteristics of ITWASs was developed [2].

Today, using various computer programs, we can simulate almost any accelerating sections (see, for example, [3,4]). However, the complexity of the results obtained, their strong dependence on the grid parameters and impossibility of using approximate analysis still make the development and use of semi-analytical approaches actual.

Two approximate approaches were mainly used to describe ITWASs: a coupled cavity model (see, for example, [5,6,7,8,9,10,11,12,13,14,15,16,17]) and a waveguide approximation (see, for example, [18,19,20,21,22,23,24,25]). While the first approach is based on the strict physical and mathematical foundation, the necessities of use many eigen modes, difficulties of coupling coefficient calculation and taking into account the losses in walls made the definition of parameters of coupled cavity models very approximate. Nevertheless, these models were useful in practice and together with developer skills gave good results.

The second approach is based on assumption that there are such slow parameter changes under which there are no practical differences between equations of homogeneous and inhomogeneous waveguides. Under such assumption, we can transform the definition for homogeneous waveguide $R_{ser} = E_0^2/P$ ($R_{ser}$ - serious impedance, $E_0$ - amplitude of the principal space harmonic) into an equation $R_{ser}(z) = E_0^2(z)/P(z)$, which is the base of this approach. It was a useful assumption, but nobody knows accuracy of the obtained results.

Smooth approximate models are widely used, especially in the study of beam current loading and transient effects, but so far the notion of spatial averaged electric field in the model equations (together with model equations) has not yet been rigorously defined.

Approximate model equations are used with parameters that are the slow functions of coordinate. Under assumptions of these models in the considered structures there are only two independent (forward and backward) waves which characteristics slowly change along the waveguide. Evanescent wave are ignored in these models. These features arise in mathematic physics when we use asymptotic expansions. It is obvious that approximate models are based on the several first equations of the asymptotic expansion chain of the solutions of the exact equations (if such equations exist). But at what level: on the equations of the zero (Eikonal) or first (WKB) order?

The possibility of using the WKB approach to describe the ITWAS gives not only a simplification of the calculation. It also allows the use of simpler physical models of transient processes. Using the traveling wave concept simplifies the understanding of pulsed-excited ITWAS transients and the development of methods to mitigate their effect on beam parameters.

Difficulties in describing the ITWASs arise from the fact that there were not obtained closed and rigorous equations (except the Maxwell equations with boundary conditions) for parameters of the ITWAS from which we could obtain approximate models by using different mathematic methods.

There are works that study waves in slowly varying band-gap media on the base of analyses of differential operators without assumption that the wavelength is long compared with the size of the repeating cell (see, for example, [26,27,28,29,30] and the literature cited there). Results obtained in these works cannot be used for description ITWASs as there are no suitable smooth differential operators. Taking into account this circumstance it was proposed to use difference equations to describe ITWASs [31]. The first attempt was made on the base of the coupled cavities model that was developed with using many eigen modes and rigorous calculation of coupling coefficients, but without losses in the walls [16]. Obtained difference equations that connect the values of electric field in different points of resonators correctly describe the main waves but also contain different spurious oscillations. The reason of appearance of spurious oscillations and its influence on the solutions are not quite clear.

To explore other possibilities of using difference equations and approximate methods, we have proposed a simple but rigorous model of ITWAS [32]. This model is based on the method of Coupled Integral Equations (CIE) (see, for example, [33]). Using the theory of solving matrix equations (see [34,35,] and sited there literature) and the decomposition method [36], we obtained new matrix difference equations, on the basis of which various approximate approaches, including the WKB approach, can be developed.

It is worth to note that the unknowns in the matrix difference equations are vectors which components are the moments of electric fields on the surfaces that divide the chain resonators. Determining these moments gives possibility to calculate electromagnetic fields in any point of resonator. Therefore, proposed equations are not direct equations for the electric field. This circumstance makes it difficult to analyze the foundations of the equations that are currently used.

In this paper we present the results of using proposed approach to study the properties of different accelerating sections.

## 2. MODEL OF ITWAS. BASIC EQUATIONS

In this section, we present the basic equations of the model, the derivation of which is presented in the work [32]. We consider the chain of $N_R$ cylindrical resonators that couple through cylindrical openings in the thin diaphragms. End resonators through cylindrical openings are connected to the cylindrical waveguides. The resonator volumes are filled with a dielectric, the dielectric constant of which has an imaginary part $\varepsilon = 1 + i\varepsilon''$. With this choice one can take into account the losses and preserve the orthogonality of the waveguide cylindrical functions.

Electric field in the $k$-th resonator is determined as

$$E_z^{(k)}(z_k, r=0) = \left(T_1^{E(k)}(z_k)\right)^T C^{(k)} - \left(T_2^{E(k)}(z_k)\right)^T C^{(k+1)}, \quad (1)$$

where $0 < z_k < d_k$, $T_1^{E(k)}(z_k), T_2^{E(k)}(z_k), C^{(k)}$ are the complex $N_m$-dimensional vectors (see their definition in [32]). Vector $C^{(k)}$ determines the electric field on the opening of the $k$-th diaphragm

$$E_r^{(k)} = \sum_{s=1}^{N_m} C_s^{(k)} \varphi_s\left(\frac{r}{a_k}\right), \quad (2)$$

where $\varphi_s$ is a set of basic functions.

Making special decomposition [32,36][1]

$$C^{(k)} = \Xi^{(k)}\left(\tilde{C}^{(k,1)} + \tilde{C}^{(k,2)}\right), \quad (3)$$

we get the system of difference matrix equations:

$$\tilde{C}^{(k+1,1)} = \left\{M^{(k+1,1)} + \tilde{M}^{(k+1,1)}\left(M^{(k,1)} - M^{(k+1,1)}\right)\right\}\tilde{C}^{(k,1)} + $$
$$+ \tilde{M}^{(k+1,1)}\left(M^{(k,2)} - M^{(k+1,2)}\right)\tilde{C}^{(k,2)},$$
$$\tilde{C}^{(k+1,2)} = \left\{M^{(k+1,2)} + \tilde{M}^{(k+1,2)}\left(M^{(k,2)} - M^{(k+1,2)}\right)\right\}\tilde{C}^{(k,2)} + $$
$$+ \tilde{M}^{(k+1,2)}\left(M^{(k,1)} - M^{(k+1,1)}\right)\tilde{C}^{(k,1)},$$
$$(4)$$

where $k = 2,3,...,N_R - 2$, $M$ are complex $N_m \times N_m$ matrices (see their definition in [32]).

If elements of matrices $M^{(k,i)}$ vary sufficiently slowly with $k$, then the differences $\left|M_{s,m}^{(k+1,i)} - M_{s,m}^{(k,i)}\right|$ are the small values and we can neglect some of them and get:

Eikonal approximation

$$\tilde{C}^{(k+1,1)} = M^{(k,1)}\tilde{C}^{(k,1)},$$
$$\tilde{C}^{(k+1,2)} = M^{(k,2)}\tilde{C}^{(k,2)}, \quad (5)$$

WKB approximation

$$\tilde{C}^{(k+1,1)} = \tilde{\tilde{M}}^{(k+1,1)}\tilde{C}^{(k,1)} = $$
$$= \left\{M^{(k+1,1)} + \tilde{M}^{(k+1,1)}\left(M^{(k,1)} - M^{(k+1,1)}\right)\right\}\tilde{C}^{(k,1)},$$
$$\tilde{C}^{(k+1,2)} = \tilde{\tilde{M}}^{(k+1,2)}\tilde{C}^{(k,2)} = $$
$$= \left\{M^{(k+1,2)} + \tilde{M}^{(k+1,2)}\left(M^{(k,2)} - M^{(k+1,2)}\right)\right\}\tilde{C}^{(k,2)}.$$
$$(6)$$

Electric field in the $k$-th resonator can be divided into "forward" and "backward" parts:

$$E_z^{(k)} = E_z^{(k,1)} + E_z^{(k,2)} = $$
$$= \left[\left(T_1^{E(k)}\right)^T \Xi^{(k)}\tilde{C}^{(k,1)} - \left(T_2^{E(k+1)}\right)^T \Xi^{(k+1)}\tilde{C}^{(k+1,1)}\right] + \quad (7)$$
$$+ \left[\left(T_1^{E(k)}\right)^T \Xi^{(k)}\tilde{C}^{(k,2)} - \left(T_2^{E(k+1)}\right)^T \Xi^{(k+1)}\tilde{C}^{(k+1,2)}\right].$$

In the case of homogeneous waveguide ($C^{(k,2)} = 0$, travelling wave regime)

$$E_z^{(k)}(z + dk, r = 0) = $$
$$= \Omega^T(z)\left(M^{(1)}\right)^k C^{(0,1)} = \Omega^T U \Lambda^{(1)k} U^{-1} C^{(0,1)}, \quad (8)$$

where $0 < z < d$, $\Omega = T_1^E(z) - M^{(1)}T_2^E(z)$, $U$ is the matrix of eigen vectors, $\Lambda^{(1)} = diag(\lambda_1^{(1)}, \lambda_2^{(1)}, ..., \lambda_{N_m}^{(1)})$ and $\lambda_s^{(1)}$ are the solution of the characteristic equation of homogeneous waveguide (see [32]).

From (8) it follows that we take into account $N_m$ eigen waves, including the evanescent ones. Indeed, if $C^{(0,1)}$ is a superposition of $N_m$ eigen modes $C^{(0,1)} = \sum_{s=1}^{N_m} B_s U_s$, we have

$$E_z^{(k)}(z + kd, r = 0) = \Omega^T \sum_{s=1}^{N_m} B_s \lambda_s^{(1)k} U_s. \quad (9)$$

---

[1] This decomposition can be considered as generalized decomposition into "forward" and "backward" solutions

# 3 SMOOTH TRANSITION BETWEEN TWO DISK LOADED WAVEGUIDES

To demonstrate the correctness and capabilities of the proposed rigorous model and approximate approaches, let us consider the classical problem of connecting two homogeneous waveguides using a smooth transition between them.

Consider the chain of resonators in which the aperture radii and resonator radii vary as

$$a_k = \frac{a_I + a_{II}}{2} + \frac{a_I - a_{II}}{2} \frac{arctg\{\alpha(k-k_0)\}}{arctg\{\alpha(3-k_0)\}}, \quad (10)$$
$$k = 3,...,N_R - 1,$$

$$b_k = \frac{b_I + b_{II}}{2} + \frac{b_I - b_{II}}{2} \frac{arctg\{\alpha(k_0-k)\}}{arctg\{\alpha(k_0-3)\}}, \quad (11)$$
$$k = 3,...,N_R - 2,$$

where $N_R = 2k_0 - 1 = 201$, $a_I = 0.99$ cm, $b_I = 4.08896$ cm ($\beta_{g,I} = v_{g,I}/c = 0.022$) and $a_{II} = 0.65$ cm, $b_{II} = 4.03934$ cm ($\beta_{g,II} = v_{g,II}/c = 0.0065$) are the aperture radii and resonator radii of two homogeneous disk-loaded waveguides. At frequency $f = 2.856$ GHz these waveguides have phase shift per cell $\varphi_0 = 2\pi/3$. All resonators have the same length $d = 3.4989$ cm. The sizes of couplers ($a_1, b_1$ and $a_{N_R+1}, b_{N_R}$) were chosen from the condition of matching homogeneous disk-loaded waveguides with cylindrical waveguide with radius $b_w = 4.2$ cm. In this section we will consider the lossless case $\varepsilon'' = 0$. The dependences of the aperture radii and resonator radii for the considered two values of parameter $\alpha$ are presented in Fig. 1.

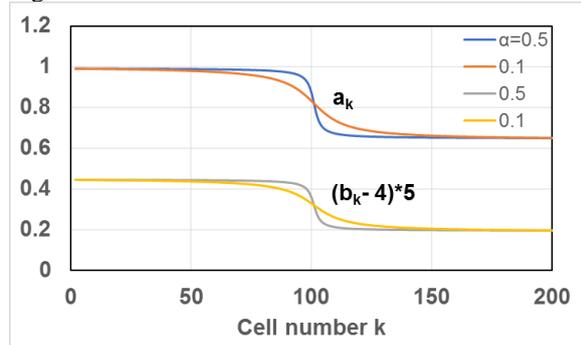

**Fig. 1**

For smooth transition ($\alpha = 0.1$) the exact and approximate (WKB) solutions practically coincide (see Fig. 2, $E^{(k)} = E_z^{(k)}(d/2 + dk, r = 0)$).

The same is true for "forward" and "backward" parts of electric field (see, Fig. 3). The detailed analysis shows that they are indeed "forward" and "backward", as their phases change in different directions.

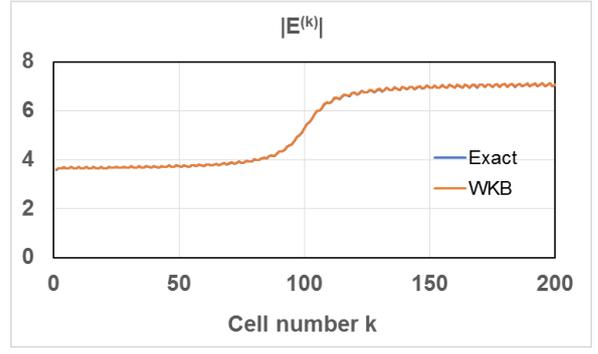

**Fig. 2**

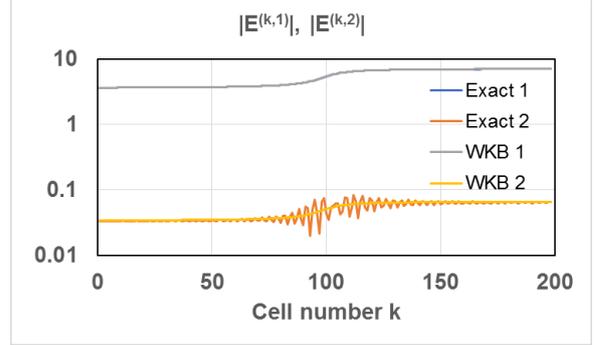

**Fig. 3**

For a steeper transition ($\alpha = 0.5$) there are difference between the exact and WKB solutions (see Fig. 4) as a reflected wave arose in the region before transition (see Fig. 5).

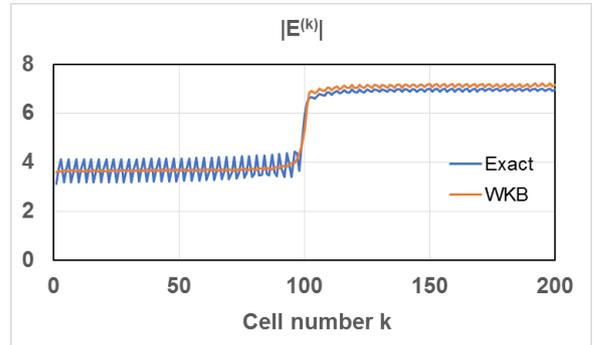

**Fig. 4**

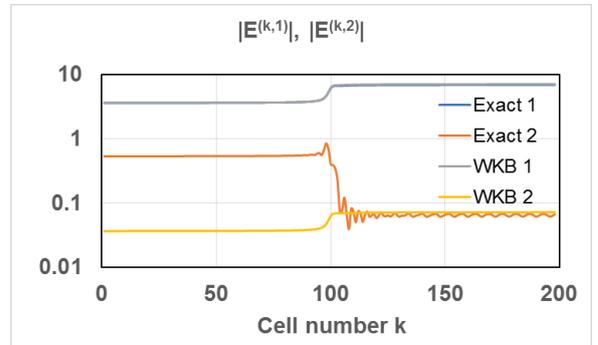

**Fig. 5**

The presented above results show that the WKB model correctly describe the considered cavity chain with a slow change in the parameters.

# 4 INHOMOGENEOUS TRAVELLING-WAVE ACCELERATING SECTIONS

The ITWASs have a unique property. Nobody knows the geometric sizes of resonators in the real section with sufficient accuracy[2]. Only the sizes of coupling holes have definite values. There are several reasons for it. The first reason is associated with the difficulty to conduct numerical modelling with an accuracy of fractions of a micrometer. The second - take into account all brazing peculiarities. And the main reason is that their knowledge does not give us useful information about the distribution of the main ITWAS characteristic - the electric field. Therefore, after preliminary selection of the resonator dimensions, fabrication and brazing, there is a tuning procedure that should give us the required electric field distribution. For today, there are two most using tuning methods: phase Ph-method and S-method. In the first method the phase shifts between resonators are tuned to the desired values by slight changing of the cavity radii (see, for example, [1]). In the second method the combinations of field meanings in some points of several cells is reduced to the desired values by the same actions [37]. For the ITWASs with phase shift $\varphi = 2\pi/3$ the tuning condition has a form

$$\left|\mathrm{Re}\left(S^{(k)}\right)\right| \Rightarrow \min,  \quad (12)$$

where

$$S^{(k)} = -\frac{\mathrm{E}^{(k-1)} + \mathrm{E}^{(k)} + \mathrm{E}^{(k+1)}}{\sqrt{3}\,\mathrm{E}^{(k)}}. \quad (13)$$

For tuned homogeneous waveguide $S^{(k)}$ have zero real parts

$$S^{(k)} = i\alpha d, \quad (14)$$

where $\alpha$ is an attenuation coefficient.

For more details of the S method and its restriction, see [38,39].

The ITWASs can be devided into three groups: with weakly [40,41,42], medium [1,43] and strong [44,45] inhomogeneity. There are sections with very fast changes (sections with a quasi-constant gradient) [46,47]. In this work we shall consider the properties of sections with medium and strong inhomogeneity.

First, we consider the properties of waveguides that can be base for the constant gradient section. It is known that for constant electric field strength the RF power must change linearly with distance [18]. A law of variation of the aperture radii was chosen so that the group velocity linearly drops from $\beta_{g,I} = =0.022$ to $\beta_{g,II} = =0.0065$ along a chain of 81 resonators. The selected values of group velocity are similar to those in the SLAC section [1]. Earlier, the results of the study of tuning methods [39] were obtained without taking into account losses of the RF field. We introduce losses by filling the resonators with media which permittivity is complex $\varepsilon = 1 + i\varepsilon''$. The value of losses ($\varepsilon'' = 10^{-4}$) was chosen from the condition of constant amplitude of the axial electric field at the resonator centers. To eliminate the influence of couplings on the calculation results, we placed 10 identical resonators before and 10 after the inhomogeneous chain (see Fig. 6).

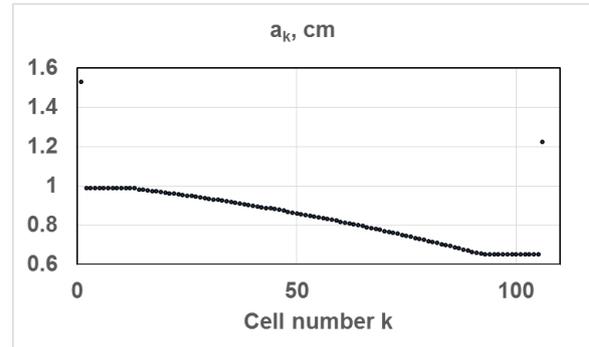

**Fig. 6**

Characteristics of axial electric field distributions (we remind that $\mathrm{E}^{(k)} = E_z^{(k)}(d/2 + dk, r = 0)$) after tuning by S-method are presented in Fig. 7 and Fig. 8. Tuning process was started from the end cells.

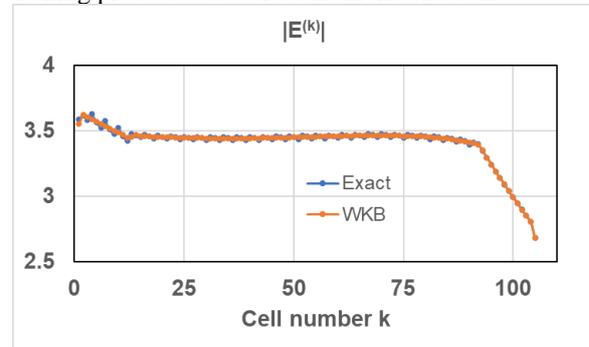

**Fig. 7 Amplitudes of the axial electric field in the middle of resonators (S method)**

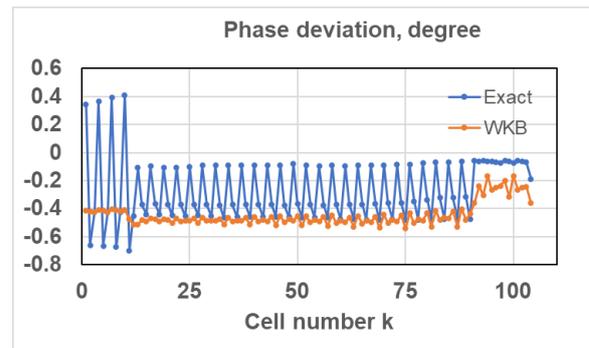

**Fig. 8 Phase deviation in the middle of resonators from the $2\pi k/3$ law (S method)**

From these results it follows that the WKB approach correctly (except for small phase deviations) describe an ITWAS that is similar to the SLAC section.

When using Ph tuning methods, we get the same amplitude distribution (Fig. 9) and slightly better phases (Fig. 10). However, WKB approach cannot be used in this case (see Fig. 9, Fig. 10), as it diverge. This divergence is not related to the presence of a turning point [48]. The reason for the divergence of the WKB method is the nonsmooth distribution of the resonator radii when using a phase tuning. Indeed, from Fig. 11 it

---

[2] Excluding sections that were assembled using diffusion bonding, requiring no tuning in the case of correct choosing of resonator frequencies

follows that there are small, but fast, oscillation of the resonator radii.

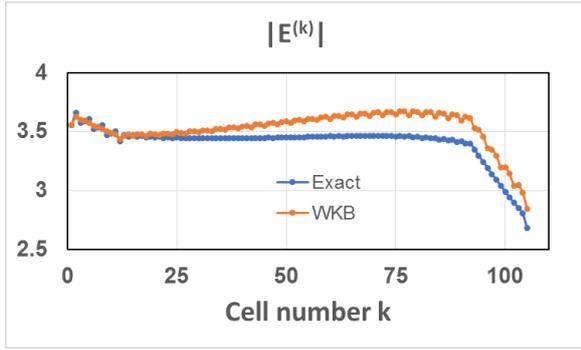

**Fig. 9 Amplitudes of the axial electric field in the middle of resonators (Ph method)**

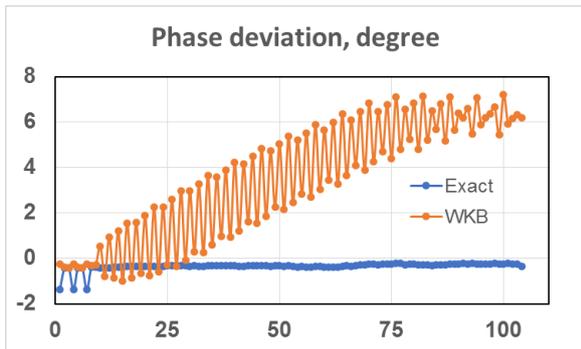

**Fig. 10 Phase deviation in the middle of resonators from the $2\pi k/3$ law (Ph method)**

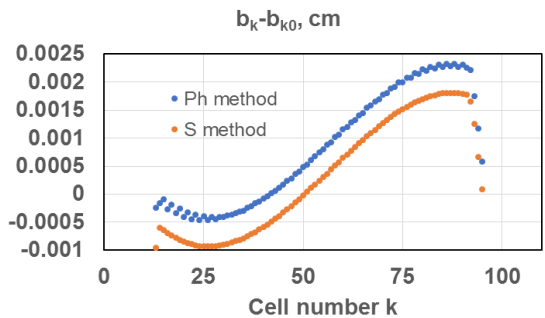

**Fig. 11 Deviation of resonator radii from some smooth distribution $b_{k,0}$ as a function of cell number. For clarity, the brown curve has been shifted down by 5 $\mu m$.**

It is interesting to note that if $\mathrm{Re}\left(S^{(k)}\right)$ change significantly when tuning (see Fig. 12), in $\mathrm{Im}\left(S^{(k)}\right)$ virtually no change occurs (see Fig. 13). This parameter has no physical meaning in nonuniform waveguides, since it follows from Fig. 13 that it can even be negative (compare with (14))

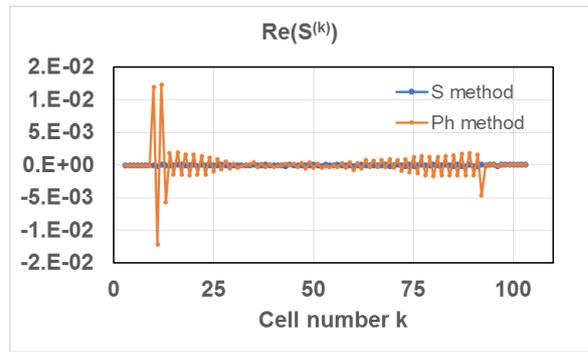

**Fig. 12**

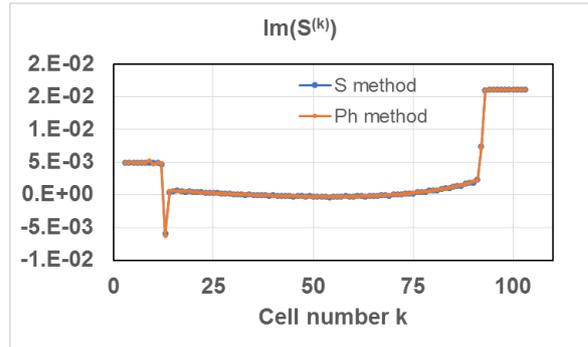

**Fig. 13**

Let's reduce the number of non-uniform cells in the chain from 81 to 31 with the same sizes of the end cells. For such waveguide S and Ph tuning methods give the same amplitude distributions (compare **Ошибка! Источник ссылки не найден.** and Fig. 16) and different phase distributions (compare Fig. 15 and Fig. 17). Ph method gives smaller phase deviations.

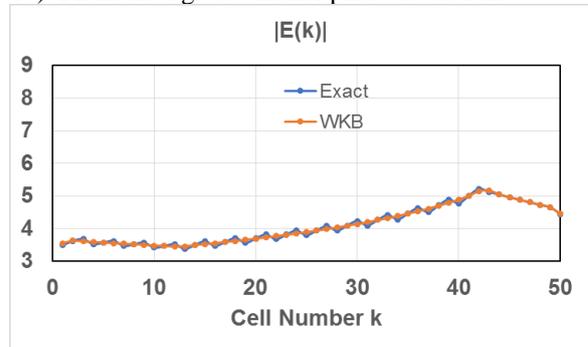

**Fig. 14 Amplitudes of the axial electric field in the middle of resonators (S method)**

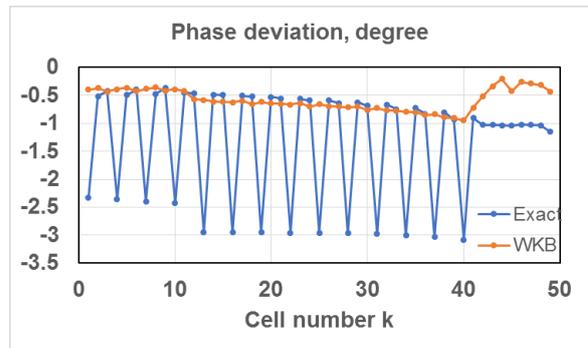

**Fig. 15 Phase deviation in the middle of resonators from the $2\pi k/3$ law (S method)**

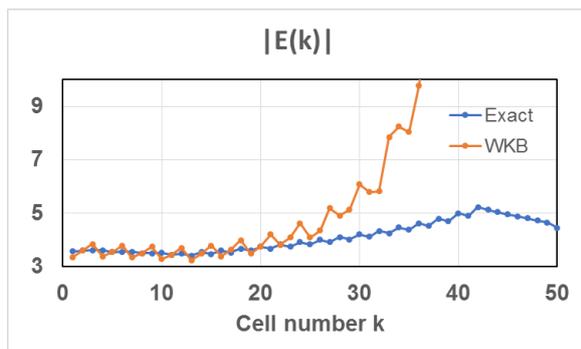

**Fig. 16** Amplitudes of the axial electric field in the middle of resonators (Ph method)

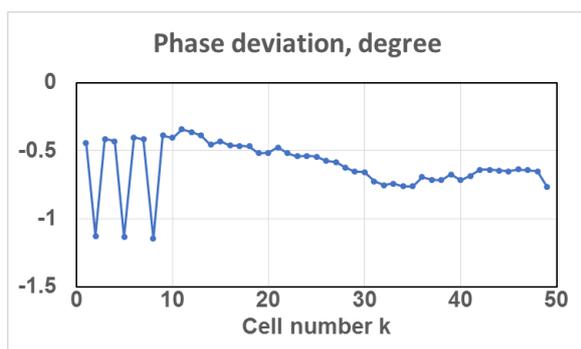

**Fig. 17** Phase deviation in the middle of resonators from the $2\pi k/3$ law (Ph method)

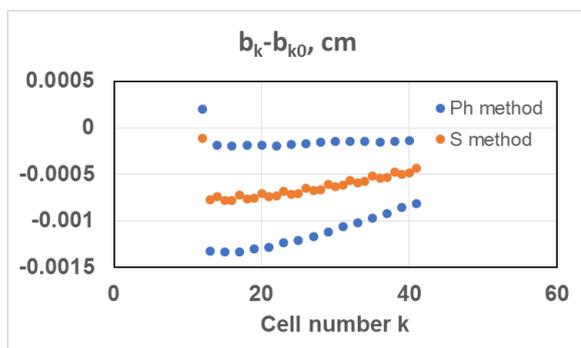

**Fig. 18** Deviation of resonator radii from some smooth distribution $b_{k,0}$

Using the WKB approach for describing such section become problematic – for the S tuning method there is discrepancy in phase distributions, for the Ph tuning method the WKB method is diverges. In this case the divergence is much stronger because the cell radii undergo larger oscillations (see Fig. 18)

## CONCLUSIONS

What does the divergence of the WKB method indicate? It indicates that the transients in this case will differ from those that arise in waveguides with smooth changes (a forward wave with a front moving with the group velocity). There will be multiple reflections from strong irregularities. For SLAC-type sections the oscillations of the resonator radii are small and the effect of multiple reflections on the transients will also be small. However, for short sections with strong inhomogeneity multiple reflections can significantly change the transients.

To avoid this, it is advisable to use tuning methods that give the necessary amplitude and phase distributions together with a smooth change of resonator frequencies.